\documentstyle[twocolumn,prl,aps,psfig,twoside,floats]{revtex}
\begin{document}
\wideabs%
{
\title{Standing and Moving Gap Solitons in Resonantly Absorbing Gratings}
\author{
Alexander~E.~Kozhekin, Gershon~Kurizki}
\address{
Chemical Physics Department, Weizmann Institute of Science, Rehovot 76100,
Israel}
\author{and
Boris~Malomed}
\address{
Department of Interdisciplinary Studies, Faculty of Engineering, Tel Aviv
University, Tel Aviv 69978, Israel}
\maketitle
\begin{abstract}
 We present hitherto unknown forms of soliton dynamics in the
 forbidden frequency gap of a Bragg reflector, modified by periodic
 layers of near-resonant two-level systems (TLS). Remarkably, even
 extremely low TLS densities create an allowed band within the
 forbidden gap. This spectrum gives rise, for {\em any\/} Bragg
 reflectivity, to a vast family of stable gap solitons, both standing
 and moving, having a unique analytic form, an {\em arbitrary\/} pulse
 area, and inelastic collision properties.  These findings suggest new
 possibilities of transmission control, noise filtering or ``dynamical
 cavities'' (self-traps) for both weak and strong signal pulses.
\end{abstract}
\draft
}

The study of light-matter interactions in periodic dielectric
structures has developed into a vast research area. At the heart of
this area is the interplay between the resonant reflections induced by
the Bragg reflector, giving rise to photonic band gaps, and their
dynamical modifications due to nonlinear light-matter
interactions. The pulsed mode of propagation in such structures
exhibits a variety of unique fundamentally and technologically
interesting regimes: nonlinear filtering, switching, and
distributed-feedback amplification \cite{Scal}. Of particular interest
are {\em gap solitons\/} (GS), i.e., moving or standing (quiescent)
self-localized pulses, whose spectrum is centered in a gap induced by
the grating. GS in Kerr-nonlinear Bragg reflectors have been
extensively analyzed \cite{Ster94} and experimentally observed
\cite{Eggl96}. Recently, GS have also been predicted in
Bragg-reflecting second-harmonic generating media \cite{Pesc97}.

This work is dedicated to a different mechanism supporting GS in
periodic media, which is based on {\em near-resonant\/} field-atom
interactions. The first step in this direction has been made in
Ref. \cite{Kozh95}, where an {\em exact\/} moving GS solution has been
found in a periodic structure composed of thin layers of resonant
two-level systems (TLS) separated by half-wavelength non-absorbing
dielectric layers, i.e., a {\em resonantly-absorbing Bragg reflector}
(RABR). In the soliton solution obtained in Ref. \cite{Kozh95}, the
combined area of the forward- and backward-propagating pulses is
$2\pi$, characteristic of self-induced transparency (SIT) solitons in
uniform media \cite{McCa}. The existence of this soliton stems from
the cooperative resonant atomic polarizability, which compensates for
the periodic modulation of the linear polarizability in the Bragg
reflector \cite{Kozh95,Mant95}.  The analysis presented in
Ref. \cite{Kozh95} leaves several open questions of fundamental and
applied importance: ({\it i\/}) Can one overcome the basic restriction
implicit in this solution, namely, that the {\em cooperative length\/}
over which a SIT pulse is formed, must be {\em shorter\/} than the
Bragg reflection length? If this restriction is essential, then the
soliton would only exist in weakly-reflecting Bragg structures, which
can hardly serve as efficient filters that block pulses other than
GS. ({\it ii\/}) Are GS admitted in a RABR for weak pulses whose area
is {\em less\/} than $2\pi $? ({\it iii\/}) Is there a quiescent
counterpart to the moving GS, which would imply {\em complete dynamic
confinement\/} of light in the RABR?  ({\it iv\/}) What is the result
of collisions between moving GS in this system?

In this work, we present answers to the above questions: (a) The RABR
supports a vast family of GS characterized by {\em two} parameters,
the soliton amplitude and velocity (analogously to GS in
Kerr-nonlinear gratings \cite{Ster94}), which exists for {\em any
ratio\/} of reflection to cooperative lengths. (b) It includes a
subfamily of quiescent or zero-velocity (ZV) solitons with an {\em
arbitrary pulse area\/}, whose {\em exact\/} analytical form
represents an essentially novel type of soliton solutions in nonlinear
optics. (c) Moving GS are found analytically in the small-amplitude
limit, where they reduce to the nonlinear-Schr\"{o}dinger (NLS)
solitons, and also numerically, as deviations from the exact subfamily
found in \cite{Kozh95} or from the exact ZV solitons. Our simulations
indicate that both ZV and moving GS are {\em stable\/}. (d) We
simulate collisions between moving GS, demonstrating that they may be
either weakly or strongly inelastic. These findings reveal a basically
new phenomenon: the multiple reflections in a grating can effectively
change the pulse coupling to the TLS, so that soliton transmission is
not restricted to pulses of $2 \pi$ area, as in ordinary SIT, and
occurs (at an appropriate frequency) for {\em any\/} ratio of
reflection to cooperative lengths.

Our starting point are the equations for the sum $\Sigma_{+}$ of the
amplitudes of the forward- and backward- propagating electromagnetic
waves, polarization $P$, and population inversion $w$, derived in
Ref. \cite{Kozh95} from the coupled Maxwell-Bloch equations for the
bidirectional propagation in RABR:
\begin{eqnarray}
\left( \frac{\partial ^2}{\partial \tau ^2}-\frac{\partial ^2}{\partial
\zeta ^2}\right) \Sigma _{+} &=&2\frac \partial {\partial \tau }P+2i\eta
P-\eta ^2\Sigma _{+},  \label{Sigma} \\
\frac \partial {\partial \tau }P &=&w\Sigma _{+}-i\delta P,  \label{P} \\
\frac \partial {\partial \tau }w &=&-\frac 12\left( P^{*}\Sigma _{+}+P\Sigma
_{+}^{*}\right) ,  \label{w}
\end{eqnarray}
where $\tau$ and $\zeta$ are the normalized time and propagation
distance, $\delta$ is the effective detuning of the field from the
atomic resonance. The key parameter $\eta=l_c/l_r$ is the ratio of the
Arrechi-Courtens cooperativity length ($l_c=2c/\omega_p$, where
$\omega_p$ is an effective plasma frequency \cite{McCa}) to the
reflection length ($l_r=4 d \epsilon_0 / \pi \Delta \epsilon$, where
$d$ is the period and $\Delta \epsilon$ the variation of the
dielectric index of the periodic structure with average dielectric
index $\epsilon_0$) \cite{Kozh95}. The crucial assumption in the
derivation of Eqs. (\ref{Sigma}-\ref{w}) is that the {\em resonant
absorbers are confined to thin layers\/}, periodically inserted
between passive thick dielectric layers. The system
(\ref{Sigma})-(\ref{w}) can be simplified: substituting into
Eq.~(\ref{w}) the expression for $\Sigma_{+}$ following from
(\ref{P}), one obtains an equation that can be explicitly integrated
to yield a simple algebraic relation
\begin{equation}
w=\pm \sqrt{1-\left| P\right| ^2},  \label{eliminate}
\end{equation}
that should be further inserted into Eq.~(\ref{Sigma}).

\begin{figure}[htb]
\centerline{ 
\begin{tabular}{cc}
\psfig{file=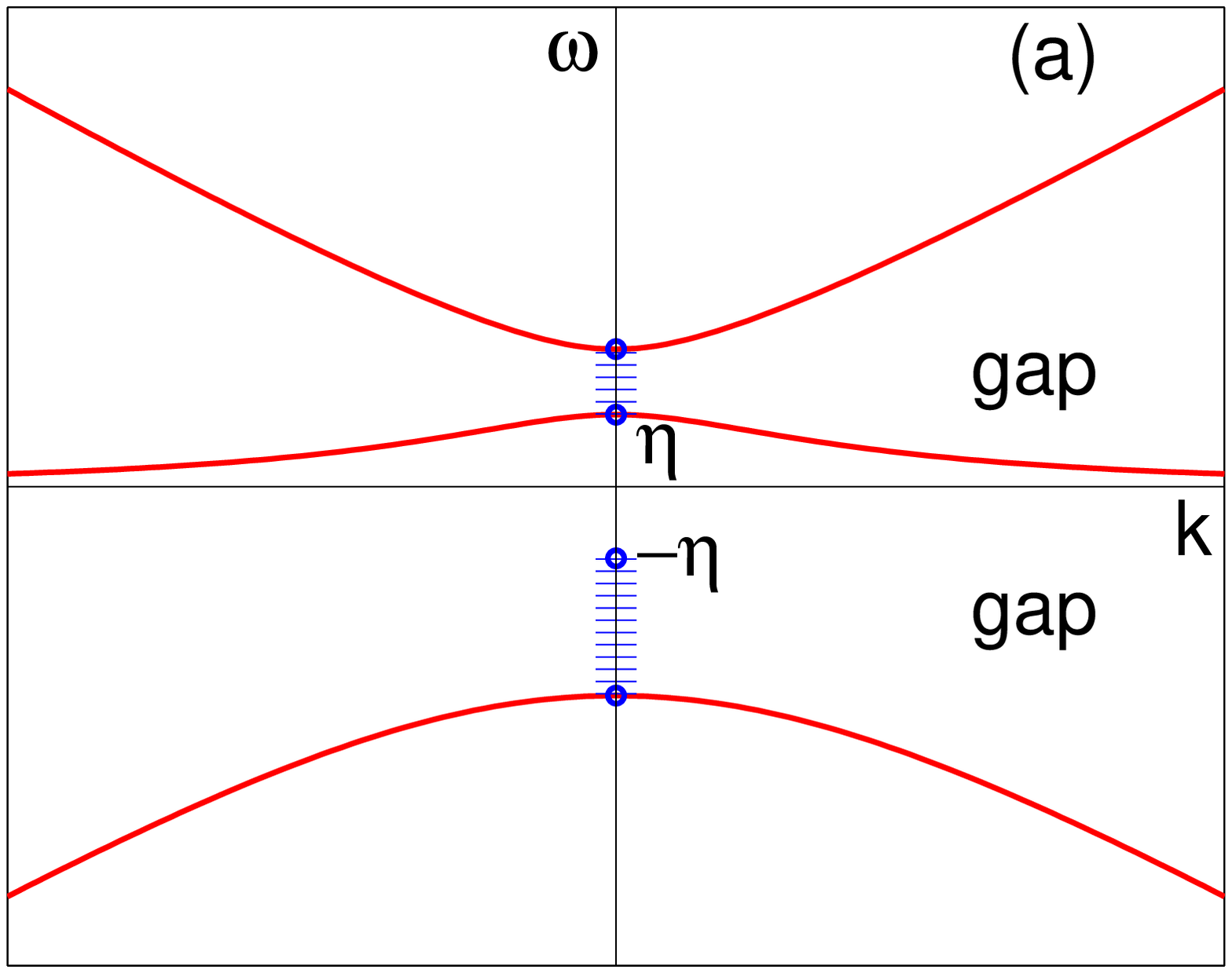,width=1.6in} &
\psfig{file=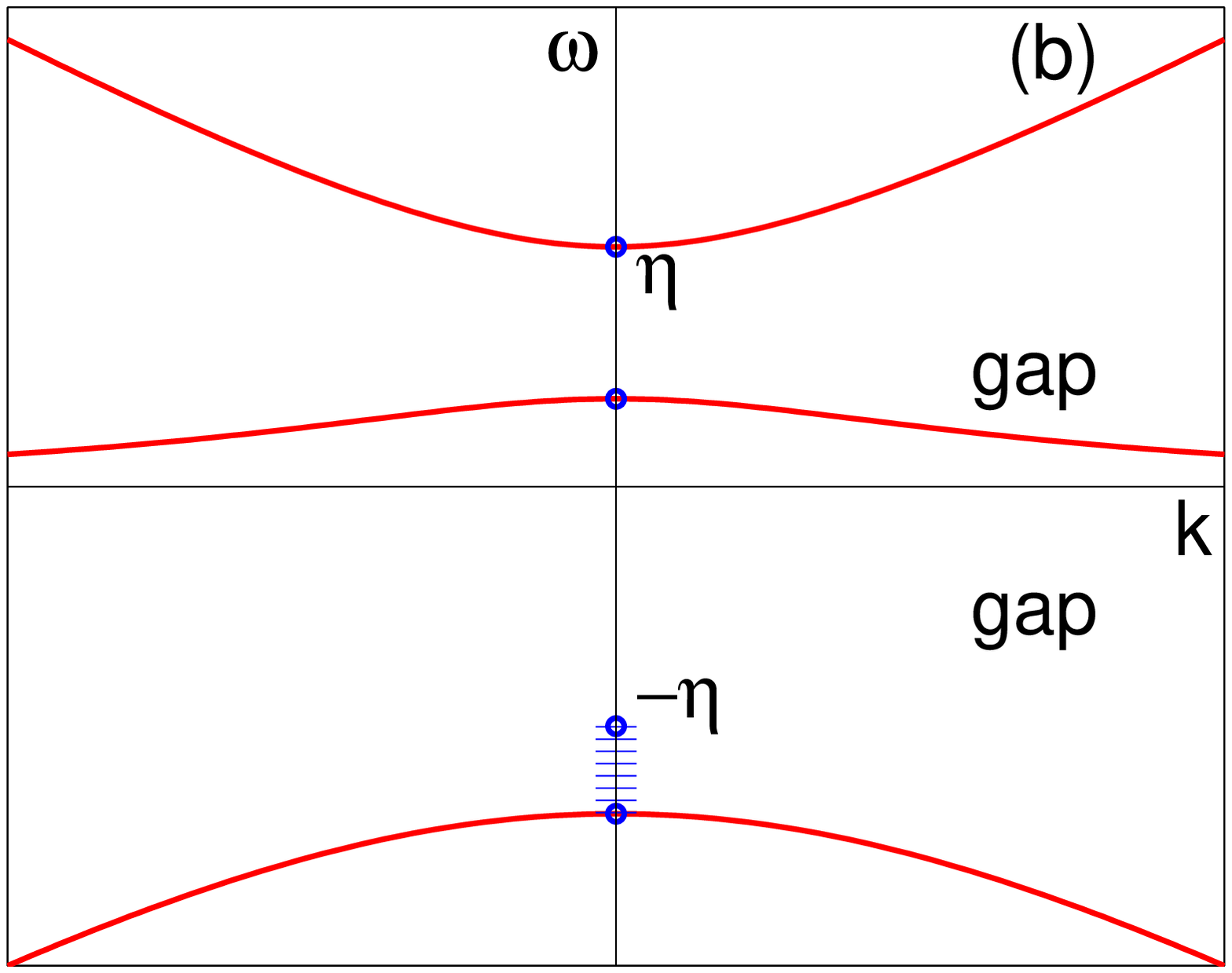,width=1.6in} 
\end{tabular}
}
\caption{The spectrum of the linearized RABR model
  (\ref{Sigma})-(\ref{w}). The parts of the upper and lower gaps
  filled with the solitons are shaded: (a) the weak-reflectivity case
  when the condition (\ref{condition}) holds ($\eta=0.6$, $\delta=0$);
  (b) the opposite case ($\eta=2$, $\delta=0$).}
\label{fig:disper}
\end{figure}

First, we linearize Eqs. (\ref{Sigma})-(\ref{w}) in order to obtain a
crucially important characteristic of the model, its dispersion
relation, by fixing $w=-1$ and substituting into the linearized
equations $\Sigma_{+}, P \sim \exp \left( ik\zeta -i\omega \tau
\right)$. The resulting dispersion relation:
\begin{equation}
\left( \omega -\delta \right) \left[ \omega ^2-k^2-\left( 2+\eta ^2\right)
\right] +2\left( \eta -\delta \right) =0,  \label{dispersion}
\end{equation}
is displayed in Fig. \ref{fig:disper}. The frequencies corresponding
to $k=0$ are $\omega =\eta $ and $\omega =-\frac 12 \left( \eta
-\delta \right) \pm \sqrt{2+\frac 14\left( \eta +\delta \right) ^2}$,
while at $k^2\rightarrow \infty $ the asymptotic expressions for
different branches of the dispersion relation are $\omega =\pm k$ and
$ \omega =\delta +2\left( \eta -\delta \right) k^{-2}$. Thus, the
linearized spectrum always splits into {\em two\/} gaps, separated by
an allowed band, except for the special case, $\eta =\eta _0\equiv
\frac 12\delta +\sqrt{1+\frac 14\delta ^2}$, when the upper gap closes
down. The upper and lower band edges are those of the periodic
structure, shifted by the induced TLS polarization in the limit of a
strong reflection. They approach the SIT spectral gap for forward- and
backward-propagating waves \cite{Mant95} in the limit of weak
reflection. The allowed middle band corresponds to a collective atomic
polarization excitation.

Inside these gaps, ZV solitons are sought in the form
\begin{equation}
\Sigma_{+} \left( \zeta ,\tau \right) =e^{-i\chi \tau } \sigma (\zeta
), \; P \left( \zeta ,\tau \right) =ie^{-i\chi \tau} q(\zeta ),
\label{soliton}
\end{equation}
with real functions $\sigma (\zeta )$ and $q(\zeta )$, $\chi$ being
the frequency detuning from the gap center. Using Eqs. (\ref{P}) and
(\ref{eliminate}), we can express the variables $w$ and $q$ in terms
of $\sigma$:
\begin{eqnarray}
w&=&\pm \left( \chi -\delta \right) \left[ \sigma ^2+\left( \chi -\delta
\right) ^2\right] ^{-1/2}, \nonumber \\
q&=&\pm \sigma \left[ \sigma ^2+\left( \chi
-\delta \right) ^2\right] ^{-1/2}.  \label{soliton_wq}
\end{eqnarray}
The remaining equation for $\sigma (\zeta )$ is 
\begin{eqnarray}
\frac{d^2\sigma }{d\varsigma ^2}&=&-\frac{dF}{d\sigma }, \quad
F(\sigma )\equiv
-\left[ \frac 12\left( \eta ^2-\chi ^2\right) \sigma ^2 \right. \nonumber
\\
&\pm& \left. 2\left( \eta
-\chi \right) \left( \sqrt{\left( \chi -\delta \right) ^2+\sigma ^2}-\left|
\chi -\delta \right| \right) \right] , \label{sigma} 
\end{eqnarray}
where the $\pm$ signs correspond to those in Eqs. (\ref{soliton_wq}).

Equation (\ref{sigma}) is the Newton equation of motion for a particle
with the coordinate $\sigma $ in the potential $F(\sigma )$.  It gives
rise to a soliton-like solution \cite{book}, provided that the upper
sign is chosen in Eqs. (\ref{soliton_wq}), and the frequency $\chi $
satisfies the conditions $|\chi |>\eta $, $|(\chi +\eta )(\chi -\delta
)|<2$. These inequalities can be solved to yield, in an explicit form,
{\em two} frequency intervals in which one has the ZV gap solitons: at
$\chi >0$, it is
\begin{equation}
\eta <\chi <\sqrt{\frac 14(\eta +\delta )^2+2}-\frac 12(\eta
-\delta ),  \label{pos_omega}
\end{equation}
which only exists for small and moderate $\eta$ (the weak
Bragg-reflectivity limit)
\begin{equation}
\eta <\frac 12\delta +\sqrt{1+\frac 14\delta ^2},  \label{condition}
\end{equation}
and another interval at $\chi <0$, which exists for any reflectivity 
\begin{equation}
\eta <-\chi <\sqrt{\frac 14(\eta +\delta )^2+2}+\frac 12(\eta -\delta ).
\label{neg_omega}
\end{equation}
On comparing these expressions with the spectrum shown in
Fig. \ref{fig:disper}, we conclude that part of the lower gap is
always empty from solitons, while the upper gap is completely filled
with ZV solitons in the weak-reflectivity case (\ref{condition})
(Fig. \ref{fig:disper}(a)), and completely empty in the
strong-reflectivity case (\ref{neg_omega})
(Fig. \ref{fig:disper}(b)). It is relevant to mention that a partly
empty gap has recently been found in the model combining Bragg
reflection and second harmonic generation \cite{Pesc97}.

Inside the frequency intervals (\ref{pos_omega}) and
(\ref{neg_omega}), Eq. (\ref{sigma}) can be integrated by means of the
substitution
\begin{equation}
\sigma (\zeta )\equiv 2\left| \chi -\delta \right| \rho (\zeta )\left(
1-\rho ^2(\zeta )\right) ^{-1}.\;  \label{rho}
\end{equation}
This yields the soliton shape in an implicit form, i.e., $\zeta$
vs. $\rho$:
\begin{eqnarray}
|\zeta | &=&\sqrt{2\left| \frac{\chi -\delta }{\chi -\eta }\right| }\left[
\left( 1-\rho _0^2\right) ^{-1/2}\tan ^{-1}\left( \sqrt{\left( \rho
_0^2-\rho ^2\right) /\left( 1-\rho_0 ^2\right) }\right) \right.  \nonumber \\
&+&\left. \left( 2\rho _0\right) ^{-1}\ln \left( (\rho _0+\sqrt{\rho
_0^2-\rho ^2})/\rho \right) \right] ,  \label{zeta}
\end{eqnarray}
where $\rho _0^2\equiv 1-\frac 12|(\chi +\eta )(\chi -\delta )|$ (note
that $\rho_0^2$ is positive under the above conditions
(\ref{pos_omega})-(\ref{neg_omega})).  It can be checked that this ZV
gap soliton is always {\em single}-humped.  Its amplitude can be found
from Eq. (\ref{zeta}),
\begin{equation}
\sigma _{\max }=4\rho _0/\sqrt{\left| \chi +\eta \right| }.
\label{amplitude}
\end{equation}

\begin{figure}[htb]
\centerline{ 
\begin{tabular}{cc}
\psfig{file=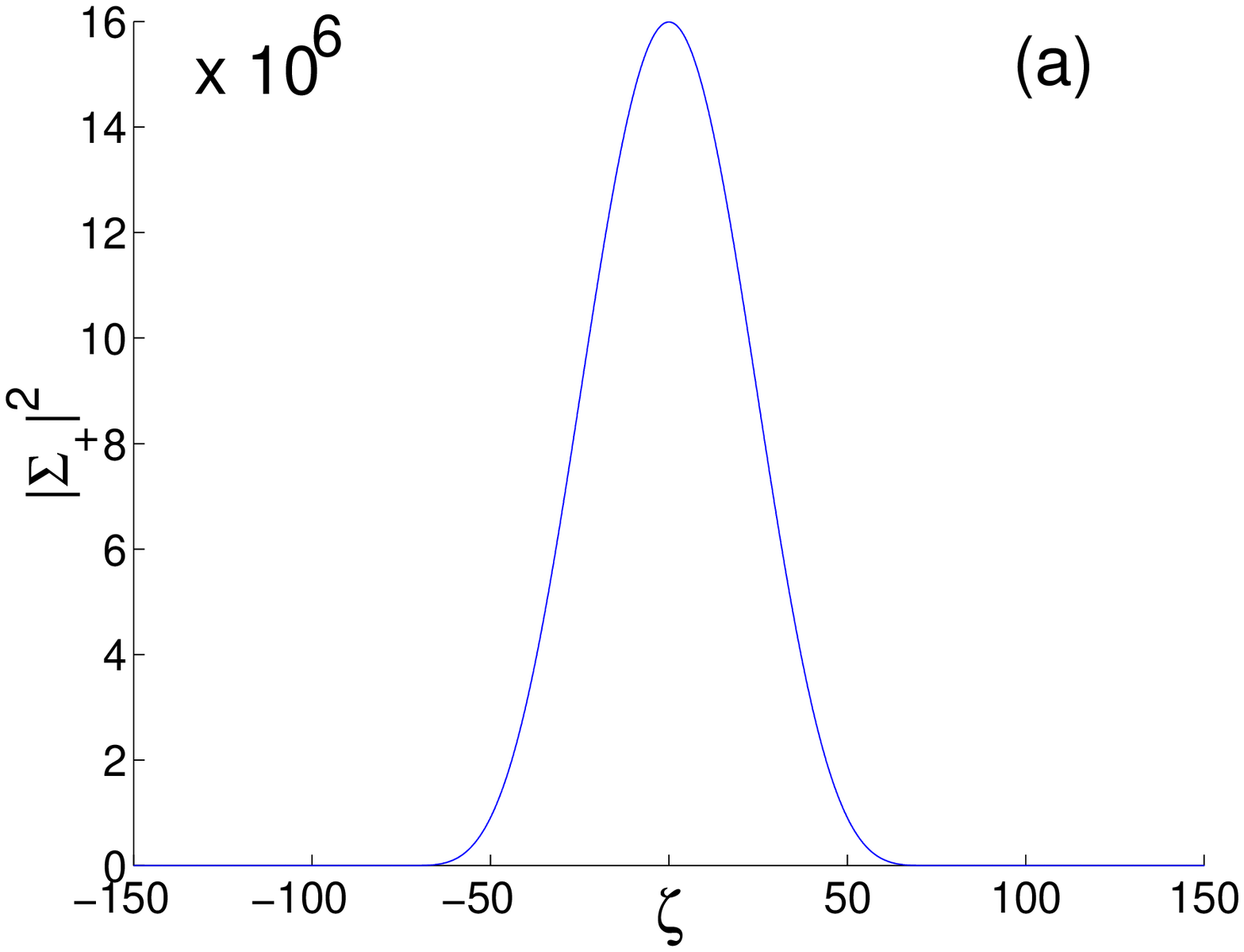,width=1.75in} &
\psfig{file=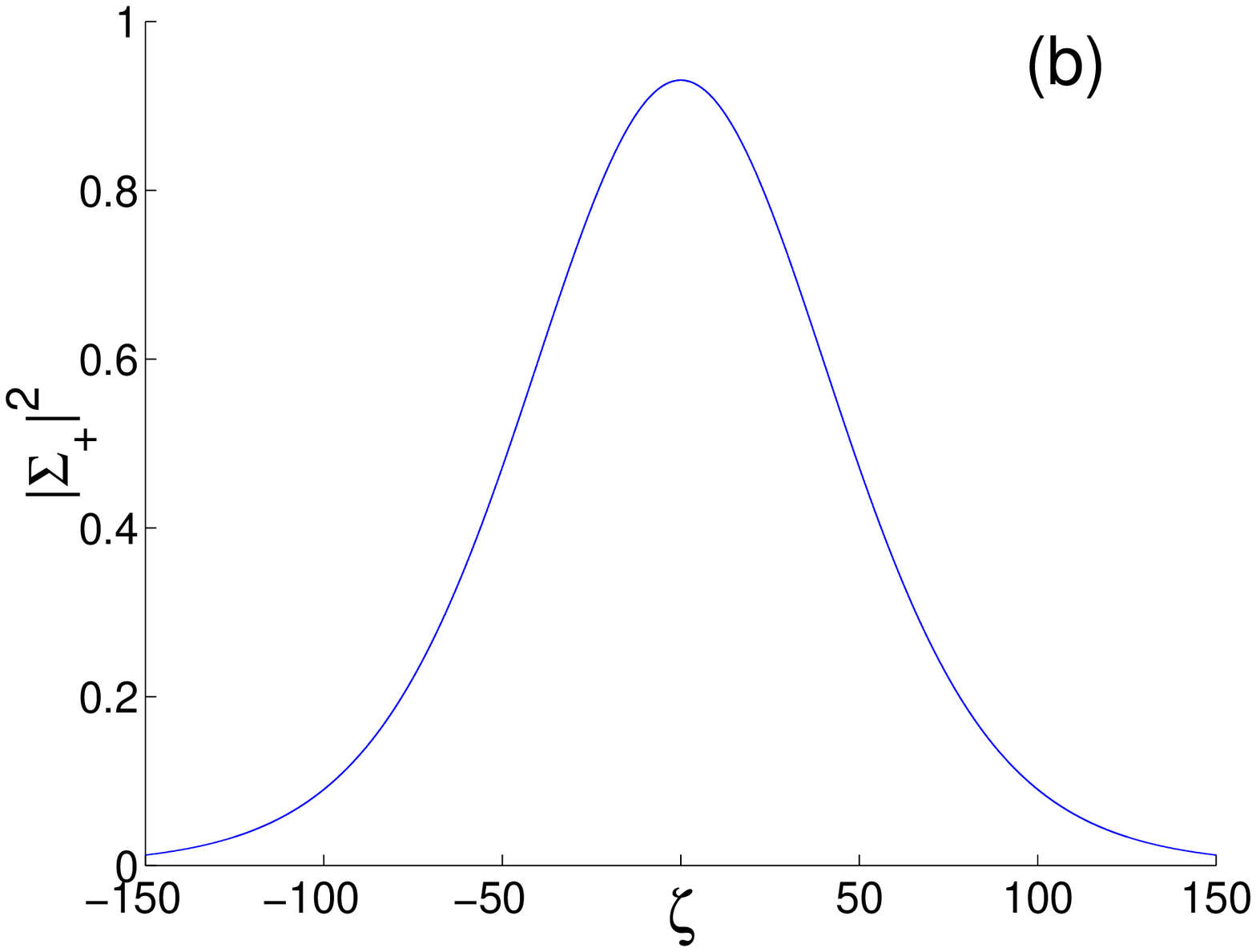,width=1.75in} \\ 
\psfig{file=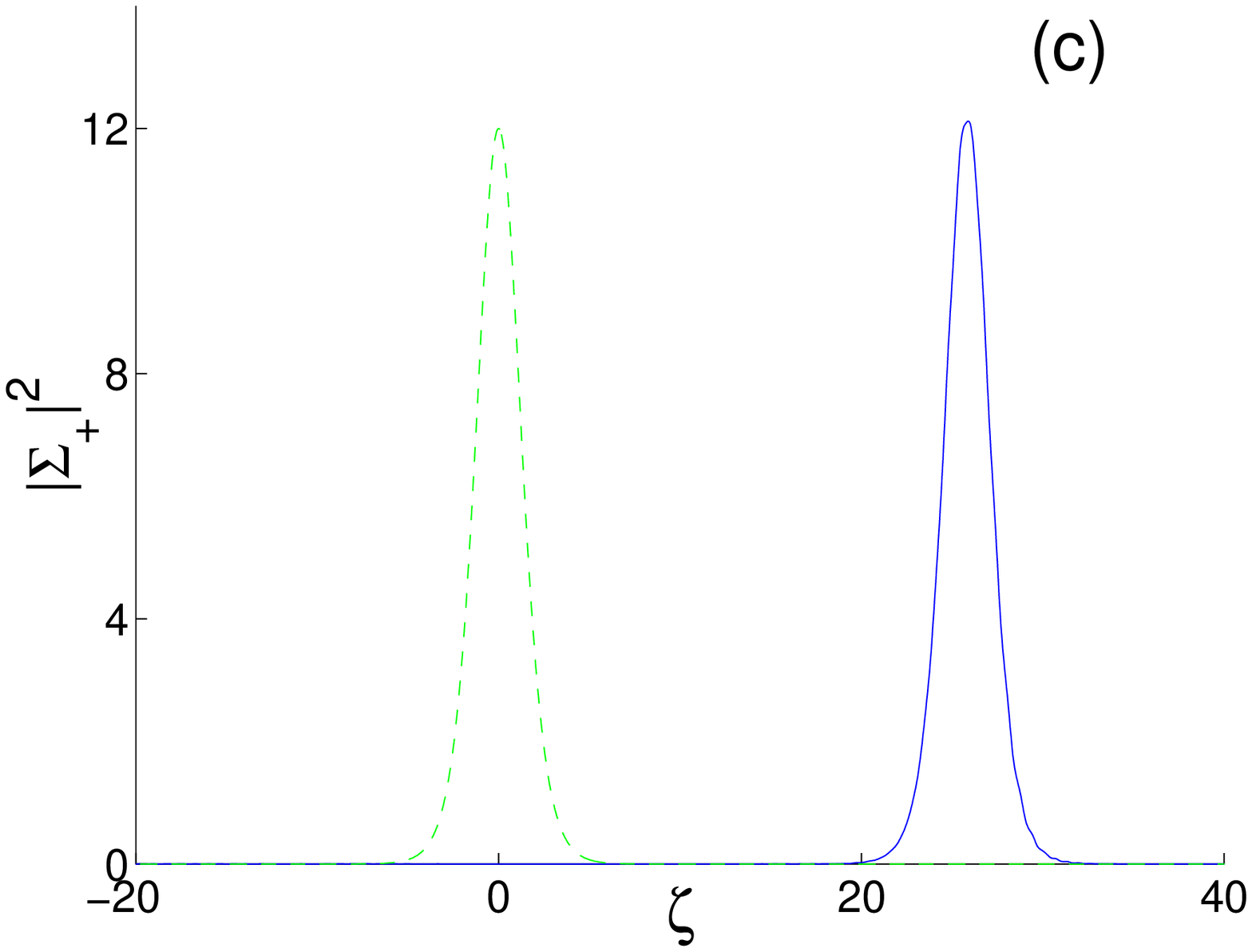,width=1.75in} &
\psfig{file=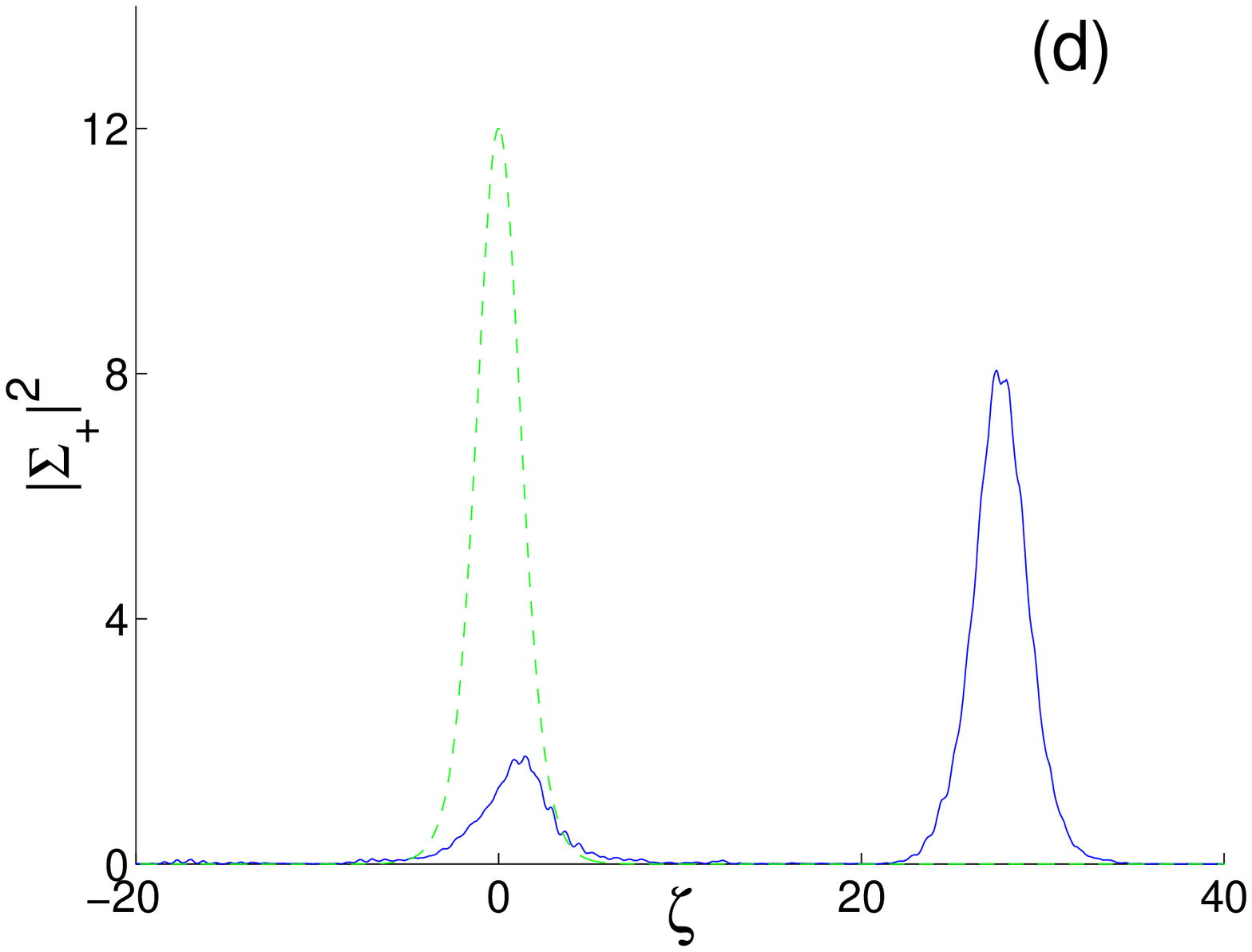,width=1.75in} 
\end{tabular}
}
\caption{Zero-velocity solitons $\left| \Sigma_{+} (\zeta )\right|^2 $
 (a) $\delta=0$, $\eta=0.9$, $\chi=-0.901$ (divergent width and
 amplitude); (b) idem, but for $\chi=0.901$ (divergent width and
 finite amplitude).  (c) Pulse(s) obtained as a result of ``pushing''
 of zero-velocity solitons (dashed lines), by the initial multiplier
 $\exp (-ip\zeta )$ after a sufficiently long evolution ($\tau=400$)
 (solid lines).  $\delta=0$, $\eta=4$, $\chi=-4.4$, and $p=0.1$. (d)
 idem, but for $p=0.5$.}
\label{fig:push}
\end{figure}

The most drastic difference of these new solitons from the well-known
SIT pulses \cite{McCa} is that the area of the ZV soliton is not
restricted to $2\pi$, but, instead, may take an {\em arbitrary}
value. As mentioned above, this basic new result shows that the Bragg
reflector can enhance (by multiple reflections) the field coupling to
the TLS, so as to make the pulse area {\em effectively\/} $2 \pi$. In
the limit of the small-amplitude and small-area solitons, $\rho_0^2
\ll 1$, Eq. (\ref{zeta}) can be easily inverted, the ZV soliton
becoming a broad {\rm sech}-like pulse\/:
\begin{equation}
\sigma \approx 2|\chi -\delta |\rho _0\,{\rm sech}\left( \sqrt{2\left|
\frac{\chi -\eta }{\chi -\delta }\right| }\rho _0\zeta \right) .
\label{small}
\end{equation}
In the opposite limit, $1-\rho _0^2\rightarrow 0$, i.e., for
vanishingly small $|\chi +\eta |$, the soliton's amplitude
(\ref{amplitude}) becomes very large, and further analysis reveals
that, in this case, the soliton is characterized by a {\em broad
central part\/} with a width $\sim \left( 1-\rho _0^2\right) ^{-1/2}$
(Fig. \ref{fig:push}(a)). Another special limit is $\chi -\eta
\rightarrow 0$. It can be checked that in this limit, the amplitude
(\ref{amplitude}) remains finite, but the {\em soliton width
diverges\/} as $|\chi -\eta |^{-1/2}$ (Fig. \ref{fig:push}(b)). Thus,
although the ZV soliton has a single hump, its shape is, in general,
strongly different from that of the traditional
nonlinear-Schr\"odinger (NLS) {\rm sech} pulse.

The stability of the ZV gap solitons was tested numerically, by means
of direct simulations of the full system (\ref{Sigma}) - (\ref{w}),
the initial condition taken as the exact soliton with a small
perturbation added to it.  Running the simulations at randomly chosen
values of the parameters, we have always found the ZV GS to be
apparently {\em stable}. However, possibility of their dynamical and
structural instability need to be further investigated, as has been
done in the case of GS in a Kerr-nonlinear fiber with a grating
\cite{Bara98,Cham98}.

Although the system of Eqs. (\ref{Sigma}) - (\ref{w}) is not
explicitly Galilean- or Lorentz-invariant, translational invariance is
expected on physical grounds. Hence, a full family of soliton
solutions should have velocity as one of its parameters. This can be
explicitly demonstrated in the limit of the small-amplitude
large-width solitons [cf. Eq. (\ref{small})]. We search for the
corresponding solutions in the form $\Sigma _{+}(\zeta ,\tau )=\sigma
(\zeta ,\tau )\exp \left( -i\omega _0\tau \right) ,\;P(\zeta ,\tau
)=iq(\zeta ,\tau )\exp \left( -i\omega _0\tau \right) $
(cf. Eqs. (\ref {soliton})), where $\omega _0$ is the frequency
corresponding to $k=0$ on any of the three branches of the dispersion
relation (\ref{dispersion}) (see Fig.  \ref{fig:disper}), and the
functions $\sigma (\zeta ,\tau )$ and $q(\zeta ,\tau )$ are assumed to
be slowly varying in comparison with $\exp \left( -i\omega _0\tau
\right)$. Under these assumptions we arrive at the following
asymptotic equation for $\sigma (\zeta ,\tau )$:
\begin{eqnarray}
\left[ 2i\frac{\omega _0(\omega _0-\delta )^2-\eta +\delta }{(\omega
_0-\delta )^2} \frac{\partial}{\partial \tau } \right. 
&+&\frac{\partial^2}{\partial \zeta ^2}+ \label{NLS} \\ \left.
\frac{\omega _0-\eta }{(\omega _0-\delta)^3} 
|\sigma |^2 \right] \sigma &=& \left( \eta ^2-\omega
_0^2+2\frac{\omega _0-\eta }{\omega _0-\delta } \right) \sigma .
\nonumber 
\end{eqnarray}
Since this equation is of the NLS form, it has the full two-parameter
family of soliton solutions, including the moving ones \cite{book}.

In order to check the existence and stability of the {\em moving}
solitons numerically, we simulated Eqs. (\ref{Sigma}) - (\ref{w}) for
an initial configuration in the form of the ZV soliton multiplied by
$\exp (ip\zeta )$ with some wavenumber $p$, in order to ``push'' the
soliton. The results demonstrate that, at sufficiently small $p$, the
``push'' indeed produces a moving stable soliton
(Fig. \ref{fig:push}(c)). However, if $p$ is large enough, the
multiplication by $\exp (ip\zeta )$ turns out to be a more violent
perturbation, splitting the initial pulse into two solitons, one
quiescent and one moving (Fig. \ref{fig:push}(d)).

A one-parameter subfamily of moving GS was found in an exact form in
Ref. \cite{Kozh95}:
\begin{equation}
\Sigma _{+}=A_0\exp \left[ i\left( \alpha \zeta -\Delta \tau \right) \right] 
{\rm sech}\left[ \frac 12A_0u\left( \zeta -u\tau \right) \right] ,
\label{moving}
\end{equation}
where $\Delta$ is the detuning from the gap center, and the squared
amplitude $A_0^2=\left( 1-u^2\right) ^{-2}\left[ 8u^2\left(
1-u^2\right) -\eta ^2\left( 1+u^2\right) ^2\right] $ is expressed in
terms of the velocity $u$ (normalized to $c$). The values of $u$ are
restricted by the condition $A_0^2>0$, which, in particular, forbids
$u=0$.  The present analysis strongly suggests, but does not
rigorously prove, that the subfamily (\ref{moving}) belongs to a far
more general two-parameter family, whose other particular
representatives are the exact ZV solitons (\ref{zeta}) and the
approximate small-amplitude solitons determined by Eq. (\ref{NLS}).

\begin{figure}[htb]
\centerline{ \psfig{file=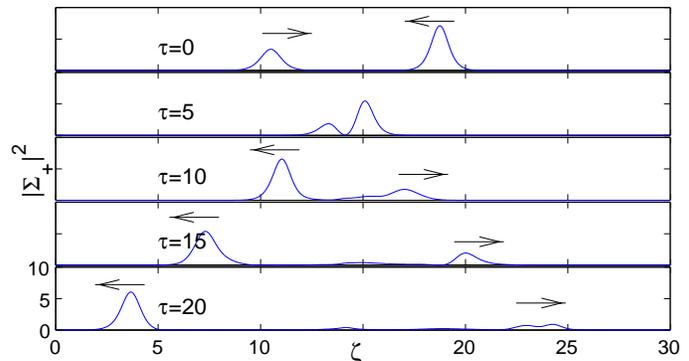,width=3.5in}}
\caption{Typical example of inelastic collisions between the solitons
 (\ref{moving}) at $\delta =0$ and $\eta =0.5$, with the velocities
 (normalized to $c$) $u_1=0.6$, $u_2=-0.75$. }
\label{fig:4}
\end{figure}

An issue of obvious interest is collisions between GS moving at
different velocities. In the asymptotic small-amplitude limit reducing
to the NLS equation (\ref{NLS}), the collision must be elastic. To get
a more general insight, we simulated collisions between two solitons
given by (\ref{moving}). The conclusion is that the collision is {\em
always inelastic\/}, directly attesting to the nonintegrability of the
model.  Typical results are displayed in Fig. \ref{fig:4}, which
demonstrates that the inelasticity may be strong, depending on the
parameters.

The present findings can be demonstrated in periodically etched
dielectric structures containing either strained quantum wells or gas
as the active TLS. For example, we can use HF gas whose active dipole
transition at the wavelength $\lambda=84$ $\mu$m is resonant with a
photonic band edge. A 4.5 mTorr gas pressure corresponds to
cooperative length $l_c=114$ cm and inhomogeneous dephasing length $c
T_2^* \sim 10^4$ cm. In a structure with Bragg reflection length
$l_r=100 \lambda \ll l_c$, we then have $\eta=-\delta=-135.7$, which
allow for GS detuned by $2\times10^3$ cm$^{-1}$ from both band edges
with $\sim 2$ ns width and $4 \times 10^9$ s$^{-1}$ Rabi
frequency. The suggested structures can be used to realize any kind of
GS: standing (ZV) GS require initial {\em localized\/} excitation of
the near-resonant medium by counterpropagating laser beams, whereas
moving GS can be launched by a single near-resonant laser beam
propagating along the structure.

In conclusion, we have demonstrated that a periodic array of
near-resonant two-level systems (TLS) combined with a Bragg grating
gives rise, for {\em any\/} Bragg reflectivity, to a vast family of
stable gap solitons, both standing and moving, having a unique
analytic form, an {\em arbitrary\/} pulse area (Fig. \ref{fig:push}),
and inelastic collision properties (Fig. \ref{fig:4}).  Remarkably,
even extremely low TLS densities create an allowed band within the
forbidden gap (Fig. \ref{fig:disper}).  These findings reveal hitherto
unknown forms of soliton dynamics in a nonintegrable, strongly
nonlinear system. Their origin is the surprisingly rich interplay
between multiple reflections and cooperative, near-resonant
field-matter interaction, which removes the pulse-area restrictions of
ordinary self-induced transparency. These findings can lead to the
realization of novel, highly advantageous filters, which can stably
transmit selected signal frequencies through their spectral gaps,
while effectively blocking others, with no restriction on the signal
pulse area. Alternatively, they can be used to spatially confine
(self-trap) light in certain frequency bands, thus creating
``dynamical cavities''.

A.K. and G.K. acknowledge the support of ISF and of the TMR network
(EU).

\begin{references}

\bibitem{Scal} M. Scalora {\it et~al.}, J. Appl. Phys. {\bf 76}, 2023
 (1994); Phys. Rev. Lett. {\bf 73}, 1368 (1994). See {\em Photonic
 Band-Gap Bibliography} compiled by J.~P.~Dowling and H.~O.~Everitt at
 WWW: {\tt \ http://hwilwww.rdec.redstone.army.mil/ \\
 MICOM/wsd/ST/RES/PBG/pbgbib.html }.

\bibitem{Ster94} D.~N. Christodoulides and R.~I. Joseph,
 Phys. Rev. Lett.  {\bf 62}, 1746 (1989); A.~B. Aceves and S. Wabnitz,
 Phys. Lett. A {\bf 141}, 37 (1989); C.~M. de~Sterke and J.~E. Sipe,
 in {\em Progress in Optics}, edited by E. Wolf (Elsevier,
 North-Holland, 1994), Vol.~{XXXIII}, Chap.~3, pp.\ 205 -- 259.

\bibitem{Eggl96}  B.~J. Eggleton {\it et~al.}, Phys. Rev. Lett. {\bf 76},
 1627 (1996).

\bibitem{Pesc97} T. Peschel, U. Peschel, F. Lederer, and
 B.~A. Malomed, Phys. Rev. E {\bf 55}, 4730 (1997); C. Conti,
 S. Trillo and G. Assanto, Phys. Rev. Lett. {\bf78}, 2341 (1997);
 Phys. Rev. E {\bf 57}, R1251 (1998); H. He and P.D.  Drummond,
 Phys. Rev. Lett. 78, 4311 (1997) .

\bibitem{Kozh95}  A. Kozhekin and G. Kurizki, Phys. Rev. Lett. {\bf 74},
 5020 (1995).

\bibitem{McCa} S.~L. McCall and E.~L. Hahn, Phys.Rev {\bf 183}, 457
 (1969); G.~L. Lamb, Jr., Rev. Mod. Phys {\bf 43}, 99 (1971).

\bibitem{Mant95} B.~I. Mantsyzov, Phys. Rev. A {\bf 51}, 4939 (1995);
 B.~I. Mantsyzov and R.~N. Kuz'min, Sov. Phys. JETP {\bf 64}, 37 (1986).

\bibitem{book} A.~C. Newell and J.~V. Moloney {\em Nonlinear
 Optics\/}, (Addison-Wesley, Redwood City CA, 1992).

\bibitem{Bara98} I.~V. Barashenkov, D.~M. Pelinovsky, and
 E.~V. Ze\-mlya\-naya, Phys. Rev. Lett. {\bf 80}, 5117 (1998).

\bibitem{Cham98}  A.~R. Champneys, B.~A. Malomed, and M.~J. Friedman, {Phys.
 Rev. Lett.} {\bf 80}, 4169 (1998).

\end{references}
\end{document}